# 1. THE COSMOLOGICAL PARAMETERS

Written August 2003 by O. Lahav (University of Cambridge) and
A.R. Liddle (University of Sussex).

## 1.1.  Parametrizing the Universe

Rapid advances in observational cosmology are leading to the
establishment of the first precision cosmological model, with many of
the key cosmological parameters determined to one or two significant
figure accuracy. Particularly prominent are measurements of cosmic
microwave anisotropies, led by the first results from the Wilkinson
Microwave Anisotropy Probe (WMAP) announced in February
2003 [1]. However the most accurate model of the Universe requires
consideration of a wide range of different types of observation, with
complementary probes providing consistency checks, lifting parameter
degeneracies, and enabling the strongest constraints to be placed.

The term 'cosmological parameters' is forever increasing in its
scope, and nowadays includes the parametrization of some functions,
as well as simple numbers describing properties of the Universe.
The original usage referred to the parameters describing the global
dynamics of the Universe, such as its expansion rate and curvature.
Also now of great interest is how the matter budget of the Universe
is built up from its constituents: baryons, photons, neutrinos, dark
matter, and dark energy. We are interested in describing the nature of
perturbations in the Universe, through global statistical descriptions
such as the matter and radiation power spectra. There may also
be parameters describing the physical state of the Universe, most
prominent being the ionization fraction as a function of time during
the era since decoupling. Typical comparisons of cosmological models
with observational data now feature about ten parameters.

### 1.1.1.  *The global description of the Universe*:

Ordinarily, the Universe is taken to be a perturbed Robertson-
Walker space-time with dynamics governed by Einstein's equations.
This is described in detail by Olive and Peacock in this volume. Using
the density parameters $\Omega_i$ for the various matter species and $\Omega_\Lambda$ for
the cosmological constant, the Friedmann equation can be written

$$\sum_i \Omega_i + \Omega_\Lambda = \frac{k}{R^2 H^2} \,, \qquad (1.1)$$

where the sum is over all the different species of matter in the
Universe. This equation applies at any epoch, but later in this article
we will use the symbols $\Omega_i$ and $\Omega_\Lambda$ to refer to the present values.
A typical collection would be baryons, photons, neutrinos, and dark
matter (given charge neutrality, the electron density is guaranteed to
be too small to be worth considering separately).

The complete present state of the homogeneous Universe can be
described by giving the present values of all the density parameters
and the present Hubble parameter $H_0 = 100h \, \mathrm{km \, s^{-1} \, Mpc^{-1}}$, and indeed one of the density parameters can be eliminated using Eq. (1.1). These also allow us to track the history of the Universe back in time, at least until an epoch where interactions allow interchanges between the densities of the different species, which is believed to have last happened at neutrino decoupling shortly before nucleosynthesis. To probe further back into the Universe's history requires assumptions about particle interactions, and perhaps about the nature of physical laws themselves.

**1.1.2.   *Neutrinos***:

The standard neutrino sector has three flavors. For neutrinos of mass in the range $5 \times 10^{-4}\,\mathrm{eV}$ to $1\,\mathrm{MeV}$, the density parameter in neutrinos is predicted to be

$$\Omega_\nu h^2 = \frac{\sum m_\nu}{94\,\mathrm{eV}}\,, \qquad (1.2)$$

where the sum is over all families with mass in that range (higher masses need a more sophisticated calculation). We use units with $c = 1$ throughout. Recent results on atmospheric and solar neutrino oscillations [2] imply non-zero mass-squared differences between the three neutrino flavors. These oscillation experiments cannot tell us the absolute neutrino masses, but within the simple assumption of a mass hierarchy suggest a lower limit of $\Omega_\nu \approx 0.001$ on the neutrino mass density parameter.

For a total mass as small as $0.1\,\mathrm{eV}$, this could have a potentially observable effect on the formation of structure, as neutrino free-streaming damps the growth of perturbations. Present cosmological observations have shown no convincing evidence of any effects from either neutrino masses or an otherwise non-standard neutrino sector, and impose quite stringent limits, which we summarize in Section 1.3.4. Consequently, the standard assumption at present is that the masses are too small to have a significant cosmological impact, but this may change in the near future.

The cosmological effect of neutrinos can also be modified if the neutrinos have decay channels, or if there is a large asymmetry in the lepton sector manifested as a different number density of neutrinos versus anti-neutrinos. This latter effect would need to be of order unity to be significant, rather than the $10^{-9}$ seen in the baryon sector, which may be in conflict with nucleosynthesis [3].

**1.1.3.   *Inflation and perturbations***:

A complete description of the Universe should include a description of deviations from homogeneity, at least in a statistical way. Indeed, some of the most powerful probes of the parameters described above come from studying the evolution of perturbations, so their study is naturally intertwined in the determination of cosmological parameters.

There are many different notations used to describe the perturbations, both in terms of the quantity used to describe the perturbations and the definition of the statistical measure. We use the dimensionless power spectrum $\Delta^2$ as defined in Olive and Peacock (also denoted $\mathcal{P}$ in some of the literature). If the perturbations obey Gaussian statistics, the power spectrum provides a complete description of their properties.

From a theoretical perspective, a useful quantity to describe the perturbations is the curvature perturbation $\mathcal{R}$, which measures the spatial curvature of a comoving slicing of the space-time. A case of particular interest is the Harrison-Zel'dovich spectrum, which corresponds to a constant spectrum $\Delta^2_{\mathcal{R}}$. More generally, one can approximate the spectrum by a power-law, writing

$$\Delta^2_{\mathcal{R}}(k) = \Delta^2_{\mathcal{R}}(k_*) \left[\frac{k}{k_*}\right]^{n-1}, \qquad (1.3)$$

where $n$ is known as the spectral index, always defined so that $n = 1$ for the Harrison-Zel'dovich spectrum, and $k_*$ is an arbitrarily chosen scale. The initial spectrum, defined at some early epoch of the Universe's history, is usually taken to have a simple form such as this power-law, and we will see that observations require $n$ close to one, which corresponds to the perturbations in the curvature being independent of scale. Subsequent evolution will modify the spectrum from its initial form.

The simplest viable mechanism for generating the observed perturbations is the inflationary cosmology, which posits a period of accelerated expansion in the Universe's early stages [4]. It is a useful working hypothesis that this is the sole mechanism for generating perturbations. Commonly, it is further assumed to be the simplest class of inflationary model, where the dynamics are equivalent to that of a single scalar field $\phi$ slowly rolling on a potential $V(\phi)$. One aim of cosmology is to verify that this simple picture can match observations, and to determine the properties of $V(\phi)$ from the observational data.

Inflation generates perturbations through the amplification of quantum fluctuations, which are stretched to astrophysical scales by the rapid expansion. The simplest models generate two types, density perturbations which come from fluctuations in the scalar field and its corresponding scalar metric perturbation, and gravitational waves which are tensor metric fluctuations. The former experience gravitational instability and lead to structure formation, while the latter can influence the cosmic microwave background anisotropies. Defining slow-roll parameters, with primes indicating derivatives with respect to the scalar field, as

$$\epsilon = \frac{m^2_{\text{Pl}}}{16\pi}\left(\frac{V'}{V}\right)^2 \quad ; \quad \eta = \frac{m^2_{\text{Pl}}}{8\pi}\frac{V''}{V}, \qquad (1.4)$$

which should satisfy $\epsilon, |\eta| \ll 1$, the spectra can be computed using the slow-roll approximation as

$$\Delta^2_{\mathcal{R}}(k) \simeq \left.\frac{8}{3m^4_{\text{Pl}}}\frac{V}{\epsilon}\right|_{k=aH};$$
$$\Delta^2_{\text{grav}}(k) \simeq \left.\frac{128}{3m^4_{\text{Pl}}}V\right|_{k=aH}. \qquad (1.5)$$

In each case, the expressions on the right-hand side are to be evaluated when the scale $k$ is equal to the Hubble radius during inflation. The symbol '$\simeq$' indicates use of the slow-roll approximation, which is expected to be accurate to a few percent or better.

From these expressions, we can compute the spectral indices

$$n \simeq 1 - 6\epsilon + 2\eta \quad ; \quad n_{\mathrm{grav}} \simeq -2\epsilon \,. \tag{1.6}$$

Another useful quantity is the ratio of the two spectra, defined by

$$r \equiv \frac{\Delta^2_{\mathrm{grav}}(k_*)}{\Delta^2_{\mathcal{R}}(k_*)} \,. \tag{1.7}$$

The literature contains a number of definitions of $r$; this convention matches that of recent versions of CMBFAST [5] and of WMAP [6], while definitions based on the relative effect on the microwave background anisotropies typically differ by tens of percent. We have

$$r \simeq 16\epsilon \simeq -8n_{\mathrm{grav}} \,, \tag{1.8}$$

which is known as the consistency equation.

In general one could consider corrections to the power-law approximation, and indeed WMAP found some low-significance evidence that this might be needed, which we discuss later. However for now we make the working assumption that the spectra can be approximated by power laws. The consistency equation shows that $r$ and $n_{\mathrm{grav}}$ are not independent parameters, and so the simplest inflation models give initial conditions described by three parameters, usually taken as $\Delta^2_{\mathcal{R}}$, $n$, and $r$, all to be evaluated at some scale $k_*$, usually the 'statistical centre' of the range explored by the data. Alternatively, one could use the parametrization $V$, $\epsilon$, and $\eta$, all evaluated at a point on the putative inflationary potential.

After the perturbations are created in the early Universe, they undergo a complex evolution up until the time they are observed in the present Universe. While the perturbations are small, this can be accurately followed using a linear theory numerical code such as CMBFAST [5]. This works right up to the present for the cosmic microwave background, but for density perturbations on small scales non-linear evolution is important and can be addressed by a variety of semi-analytical and numerical techniques. However the analysis is made, the outcome of the evolution is in principle determined by the cosmological model, and by the parameters describing the initial perturbations, and hence can be used to determine them.

Of particular interest are cosmic microwave background anisotropies. Both the total intensity and two independent polarization modes are predicted to have anisotropies. These can be described by the radiation angular power spectra $C_\ell$ as defined in the article of Scott and Smoot in this volume, and again provide a complete description if the density perturbations are Gaussian.

### 1.1.4.   *The standard cosmological model*:

We now have most of the ingredients in place to describe the cosmological model. Beyond those of the previous subsections, there is only one parameter which is essential, which is a measure of the ionization state of the Universe. The Universe is known to be highly ionized at low redshifts (otherwise radiation from distant quasars would be heavily absorbed in the ultra-violet), and the ionized electrons can scatter microwave photons altering the pattern of observed anisotropies. The most convenient parameter to describe this is the optical depth to scattering $\tau$ (i.e. the probability that a given photon scatters once); in the approximation of instantaneous and complete re-ionization, this could equivalently be described by the redshift of re-ionization $z_{\rm ion}$.

The basic set of cosmological parameters is therefore as shown in Table 1.1. The spatial curvature does not appear in the list, because it can be determined from the other parameters using Eq. (1.1). The total present matter density $\Omega_{\rm m} = \Omega_{\rm dm} + \Omega_{\rm b}$ is usually used in place of the dark matter density.

**Table 1.1:** The basic set of cosmological parameters. We give values as obtained using particular fit to a dataset known as WMAPext+2dF, described later. We cannot stress too much that the exact values and uncertainties depend on both the precise datasets used and the choice of parameters allowed to vary, and the effects of varying some assumptions will be shown later in Table 1.2. Limits on the cosmological constant depend on whether the Universe is assumed flat, while there is no established convention for specifying the density perturbation amplitude. Uncertainties are one-sigma/68% confidence unless otherwise stated.

| Parameter | Symbol | Value |
|---|---|---|
| Hubble parameter | $h$ | $0.73 \pm 0.03$ |
| Total matter density | $\Omega_{\rm m}$ | $\Omega_{\rm m} h^2 = 0.134 \pm 0.006$ |
| Baryon density | $\Omega_{\rm b}$ | $\Omega_{\rm b} h^2 = 0.023 \pm 0.001$ |
| Cosmological constant | $\Omega_{\Lambda}$ | See Ref. 7 |
| Radiation density | $\Omega_{\rm r}$ | $\Omega_{\rm r} h^2 = 2.47 \times 10^{-5}$ |
| Neutrino density | $\Omega_{\nu}$ | See Sec. 1.1.2 |
| Density perturbation amplitude | $\Delta_{\mathcal{R}}^2(k_*)$ | See Ref. 7 |
| Density perturbation spectral index | $n$ | $n = 0.97 \pm 0.03$ |
| Tensor to scalar ratio | $r$ | $r < 0.53$ (95% conf) |
| Ionization optical depth | $\tau$ | $\tau = 0.15 \pm 0.07$ |

As described in Sec. 1.4, models based on these ten parameters are able to give a good fit to the complete set of high-quality data available at present, and indeed some simplification is possible. Observations are consistent with spatial flatness, and indeed the inflation models so far described automatically generate spatial flatness, so we can set $k = 0$; the density parameters then must sum to one, and so one can be eliminated. The neutrino energy density is often not taken as an independent parameter. Provided the neutrino sector has the standard interactions the neutrino energy density while relativistic can be related to the photon density using thermal physics arguments, and it is currently difficult to see the effect of the neutrino mass although observations of large-scale structure have already placed interesting upper limits. This reduces the standard parameter set to eight. In addition, there is no observational evidence for the existence of tensor perturbations (though the upper limits are quite weak), and so $r$ could be set to zero.* This leaves seven parameters, which is the smallest set that can usefully be compared to the present cosmological data set. This model is referred to by various names, including ΛCDM, the concordance cosmology, and the standard cosmological model.

Of these parameters, only $\Omega_r$ is accurately measured directly. The radiation density is dominated by the energy in the cosmic microwave background, and the COBE FIRAS experiment has determined its temperature to be $T = 2.725 \pm 0.001$ Kelvin [8],   corresponding to $\Omega_r = 2.47 \times 10^{-5} h^{-2}$.

In addition to this minimal set, there is a range of other parameters which might prove important in future as the dataset further improves, but for which there is so far no direct evidence, allowing them to be set to a specific value. We discuss various speculative options in the next section. For completeness at this point, we mention one other interesting parameter, the helium fraction, which is a non-zero parameter that can affect the microwave anisotropies at a subtle level. Presently, big-bang nucleosynthesis provides the best measurement of this parameter, and it is usually fixed in microwave anisotropy studies, but the data are just reaching a level where allowing its variation may become mandatory.

### 1.1.5.   *Derived parameters*:

The parameter list of the previous subsection is sufficient to give a complete description of cosmological models which agree with observational data. However, it is not a unique parametrization, and one could instead use parameters derived from that basic set. Parameters which can be derived from the set given above include the age of the Universe, the present horizon distance, the present microwave background and neutrino background temperatures, the epoch of matter-radiation equality, the epochs of recombination and decoupling, the epoch of transition to an accelerating Universe, the baryon-to-photon ratio, and the baryon to dark matter density ratio. The physical densities of the matter components, $\Omega_i h^2$, are often more useful than the density parameters. The density perturbation amplitude can be specified in many different ways other than the large-scale primordial amplitude, for instance, in terms of its effect on the cosmic microwave background, or by specifying a short-scale quantity, a common choice being the present linear-theory mass dispersion on a scale of $8 \, h^{-1}\mathrm{Mpc}$, known as $\sigma_8$.

---

* More controversially, one could argue that as no evidence against the Harrison-Zel'dovich spectrum has yet been seen, then $n$ could be set to one. We will however allow it to vary.

Different types of observation are sensitive to different subsets of the full cosmological parameter set, and some are more naturally interpreted in terms of some of the derived parameters of this subsection than on the original base parameter set. In particular, most types of observation feature degeneracies whereby they are unable to separate the effects of simultaneously varying several of the base parameters, an example being the angular diameter/physical density degeneracy of cosmic microwave anisotropies.

## 1.2. Extensions to the standard model

This section discusses some ways in which the standard model could be extended. At present, there is no positive evidence in favor of any of these possibilities, which are becoming increasingly constrained by the data, though there always remains the possibility of trace effects at a level below present observational capability.

### 1.2.1. *More general perturbations*:

The standard cosmology assumes adiabatic, Gaussian perturbations. Adiabaticity means that all types of material in the Universe share a common perturbation, so that if the space-time is foliated by constant-density hypersurfaces, then all fluids and fields are homogeneous on those slices, with the perturbations completely described by the variation of the spatial curvature of the slices. Gaussianity means that the initial perturbations obey Gaussian statistics, with the amplitudes of waves of different wavenumbers being randomly drawn from a Gaussian distribution of width given by the power spectrum. Note that gravitational instability generates non-Gaussianity; in this context, Gaussianity refers to a property of the initial perturbations before they evolve significantly.

The simplest inflation models based on one dynamical field predict adiabatic fluctuations and a level of non-Gaussianity which is too small to be detected by any experiment so far conceived. For present data, the primordial spectra are usually assumed to be power laws.

#### 1.2.1.1. *Non-power-law spectra:*

For typical inflation models, it is an approximation to take the spectra as power laws, albeit usually a good one. As data quality improves, one might expect this approximation to come under pressure, requiring a more accurate description of the initial spectra, particularly for the density perturbations. In general, one can write a Taylor expansion of $\ln \Delta_{\mathcal{R}}^2$ as

$$\ln \Delta_{\mathcal{R}}^2 \left(k\right) = \ln \Delta_{\mathcal{R}}^2 \left(k_*\right) + \left(n_* - 1\right) \ln \frac{k}{k_*} + \frac{1}{2} \frac{dn}{d\ln k}\bigg|_* \ln^2 \frac{k}{k_*} + \cdots , \quad (1.9)$$

where the coefficients are all evaluated at some scale $k_*$. The term $dn/d\ln k|_*$ is often called the running of the spectral index [9], and has recently become topical due to a possible low-significance detection by WMAP. Once non-power-law spectra are allowed, it is necessary to specify the scale $k_*$ at which quantities such as the spectral index are defined.

**1.2.1.2.** *Isocurvature perturbations:*

An isocurvature perturbation is one which leaves the total density unperturbed, while perturbing the relative amounts of different materials. If the Universe contains $N$ fluids, there is one growing adiabatic mode and $N - 1$ growing isocurvature modes. These can be excited, for example, in inflationary models where there are two or more fields which acquire dynamically important perturbations. If one field decays to form normal matter, while the second survives to become the dark matter, this will generate a cold dark matter isocurvature perturbation.

In general there are also correlations between the different modes, and so the full set of perturbations is described by a matrix giving the spectra and their correlations. Constraining such a general construct is challenging, though constraints on individual modes are beginning to become meaningful, with no evidence that any other than the adiabatic mode must be non-zero.

**1.2.1.3.** *Non-Gaussianity:*

Multi-field inflation models can also generate primordial non-Gaussianity. The extra fields can either be in the same sector of the underlying theory as the inflaton, or completely separate, an interesting example of the latter being the curvaton model [10]. Current upper limits on non-Gaussianity are becoming stringent, but there remains much scope to push down those limits and perhaps reveal trace non-Gaussianity in the data. If non-Gaussianity is observed, its nature may favor an inflationary origin, or a different one such as topological defects. A plausible possibility is non-Gaussianity caused by defects forming in a phase transition which ended inflation.

**1.2.2.** *Dark matter properties*:

Dark matter properties are discussed in the article by Drees and Gerbier in this volume. The simplest assumption concerning the dark matter is that it has no significant interactions with other matter, and that its particles have a negligible velocity. Such dark matter is described as 'cold,' and candidates include the lightest supersymmetric particle, the axion, and primordial black holes. As far as astrophysicists are concerned, a complete specification of the relevant cold dark matter properties is given by the density parameter $\Omega_{\rm cdm}$, though those seeking to directly detect it are as interested in its interaction properties.

Cold dark matter is the standard assumption and gives an excellent fit to observations, except possibly on the shortest scales where there remains some controversy concerning the structure of dwarf galaxies and possible substructure in galaxy halos. For all the dark matter to have a large velocity dispersion, so-called hot dark matter, has long been excluded as it does not permit galaxies to form; for thermal relics the mass must be above about 1 keV to satisfy this constraint, though relics produced non-thermally, such as the axion, need not obey this limit. However, there remains the possibility that further parameters might need to be introduced to describe dark matter properties relevant to astrophysical observations. Suggestions which have been made include a modest velocity dispersion (warm dark matter) and dark matter self-interactions. There remains the possibility that the dark matter comprises two separate components, *e.g.*, a cold one and a hot one, an example being if massive neutrinos have a non-negligible effect.

### 1.2.3. *Dark energy*:

While the standard cosmological model given above features a cosmological constant, in order to explain observations indicating that the Universe is presently accelerating, further possibilities exist under the general heading dark energy.[†] A particularly attractive possibility (usually called quintessence, though that word is used with various different meanings in the literature) is that a scalar field is responsible, with the mechanism mimicking that of early Universe inflation [11]. As described by Olive and Peacock, a fairly model-independent description of dark energy can be given just using the equation of state parameter $w$, with $w = -1$ corresponding to a cosmological constant. In general, the function $w$ could vary with redshift, though practical experiments devised so far would be sensitive primarily to some average value weighted over recent epochs. For high-precision predictions of microwave background anisotropies, it is better to use a scalar-field description in order to have a self-consistent evolution of the 'sound speed' associated with the dark energy perturbations.

Present observations are consistent with a cosmological constant, but it is quite common to see $w$ kept as a free parameter to be added to the set described in the previous section. Most, but not all, researchers assume the weak energy condition $w \geq -1$. In the future it may be necessary to use a more sophisticated parametrization of the dark energy.

### 1.2.4. *Complex ionization history*:

The full ionization history of the Universe is given by specifying the ionization fraction as a function of redshift $z$. The simplest scenario takes the ionization to be zero from recombination up to some redshift $z_{\mathrm{ion}}$, at which point the Universe instantaneously re-ionizes completely. In that case, there is a one-to-one correspondence between $\tau$ and $z_{\mathrm{ion}}$ (that relation, however, also depending on other cosmological parameters).

---

[†] Unfortunately this is rather a misnomer, as it is the negative pressure of this material, rather than its energy, that is responsible for giving the acceleration.

While simple models of the re-ionization process suggest that rapid ionization is a good approximation, observational evidence is mixed, as it is difficult to reconcile the high optical depth inferred from the microwave background with absorption seen in some high-redshift quasar systems, and also perhaps with the temperature of the intergalactic medium at $z \simeq 3$. Accordingly, a more complex ionization history may need to be considered, and perhaps separate histories for hydrogen and helium, which will necessitate new parameters. Additionally, high-precision microwave anisotropy experiments may require consideration of the level of residual ionization left after recombination, which in principle is computable from the other cosmological parameters.

### 1.2.5.  *Varying 'constants'*:

Variation of the fundamental constants of nature over cosmological times is another possible enhancement of the standard cosmology. There is a long history of study of variation of the gravitational constant $G$, and more recently attention has been drawn to the possibility of small fractional variations in the fine-structure constant. There is presently no observational evidence for the former, which is tightly constrained by a variety of measurements. Evidence for the latter has been claimed from studies of spectral line shifts in quasar spectra at redshifts of order two [12], but this is presently controversial and in need of further observational study.

### 1.2.6.  *Cosmic topology*:

The usual hypothesis is that the Universe has the simplest topology consistent with its geometry, for example that a flat Universe extends forever. Observations cannot tell us whether that is true, but they can test the possibility of a non-trivial topology on scales up to roughly the present Hubble scale. Extra parameters would be needed to specify both the type and scale of the topology, for example, a cuboidal topology would need specification of the three principal axis lengths. At present, there is no direct evidence for cosmic topology, though the low values of the observed cosmic microwave quadrupole and octupole have been cited as a possible signature.

## 1.3.  Probes

The goal of the observational cosmologist is to utilize astronomical objects to derive cosmological parameters. The transformation from the observables to the key parameters usually involves many assumptions about the nature of the objects, as well as about the nature of the dark matter. Below we outline the physical processes involved in each probe, and the main recent results. The first two subsections concern probes of the homogeneous Universe, while the remainder consider constraints from perturbations.

### 1.3.1. *Direct measures of the Hubble constant*:

In 1929 Edwin Hubble discovered the law of expansion of the Universe by measuring distances to nearby galaxies. The slope of the relation between the distance and recession velocity is defined to be the Hubble constant $H_0$. Astronomers argued for decades on the systematic uncertainties in various methods and derived values over the wide range, $40 \, \mathrm{km \, s^{-1} \, Mpc^{-1}} \lesssim H_0 \lesssim 100 \, \mathrm{km \, s^{-1} \, Mpc^{-1}}$.

One of the most reliable results on the Hubble constant comes from the Hubble Space Telescope Key Project [13]. The group has used the empirical period-luminosity relations for Cepheid variable stars to obtain distances to 31 galaxies, and calibrated a number of secondary distance indicators (Type Ia Supernovae, Tully-Fisher, surface brightness fluctuations and Type II Supernovae) measured over distances of 400 to 600 Mpc. They estimated $H_0 = 72 \pm 3$ (statistical) $\pm 7$ (systematic) $\mathrm{km \, s^{-1} \, Mpc^{-1}}$.[‡] The major sources of uncertainty in this result are due to the metallicity of the Cepheids and the distance to the fiducial nearby galaxy (called the Large Magellanic Cloud) to which all Cepheid distances are measured relative to. Nevertheless, it is remarkable that this result is in such good agreement with the result derived from the WMAP CMB and large-scale structure measurements (see Table 1.2).

### 1.3.2. *Supernovae as cosmological probes*:

The relation between observed flux and the intrinsic luminosity of an object depends on the luminosity distance $d_L$, which in turn depends on cosmological parameters. More specifically

$$d_L = (1 + z) \, r_e \, (z) \ , \tag{1.10}$$

where $r_e(z)$ is the coordinate distance. For example, in a flat Universe

$$r_e \, (z) = \int_0^z dz'/H \, (z') \ . \tag{1.11}$$

For a general dark energy equation of state $w(z) = p_Q(z)/\rho_Q(z)$, the Hubble parameter is, still considering only the flat case,

$$H^2 \, (z) \, / H_0^2 = (1 + z)^3 \, \Omega_{\mathrm{m}} + \Omega_Q \exp \left[ 3 X \, (z) \right] \ , \tag{1.12}$$

where

$$X \, (z) = \int_0^z \left[ 1 + w \, (z') \right] \left( 1 + z' \right)^{-1} dz', \tag{1.13}$$

and $\Omega_{\mathrm{m}}$ and $\Omega_Q$ are the present density parameters of matter and dark energy components. If a general equation of state is allowed, then one has to solve for $w(z)$ (parameterized, for example, as $w(z) = w = $ constant, or $w(z) = w_0 + w_1 z$) as well as for $\Omega_Q$.

---

[‡] Unless stated otherwise, all quoted uncertainties in this article are one-sigma/68% confidence. It is common for cosmological parameters to have significantly non-Gaussian error distributions.

Empirically, the peak luminosity of supernova of Type Ia (SNe Ia) can be used as an efficient distance indicator (*e.g.*, Ref. 14). The favorite theoretical explanation for SNe Ia is the thermonuclear disruption of carbon-oxygen white dwarfs. Although not perfect 'standard candles', it has been demonstrated that by correcting for a relation between the light curve shape and the luminosity at maximum brightness, the dispersion of the measured luminosities can be greatly reduced. There are several possible systematic effects which may affect the accuracy of the SNe Ia as distance indicators, for example, evolution with redshift and interstellar extinction in the host galaxy and in the Milky Way, but there is no indication that any of these effects are significant for the cosmological constraints.

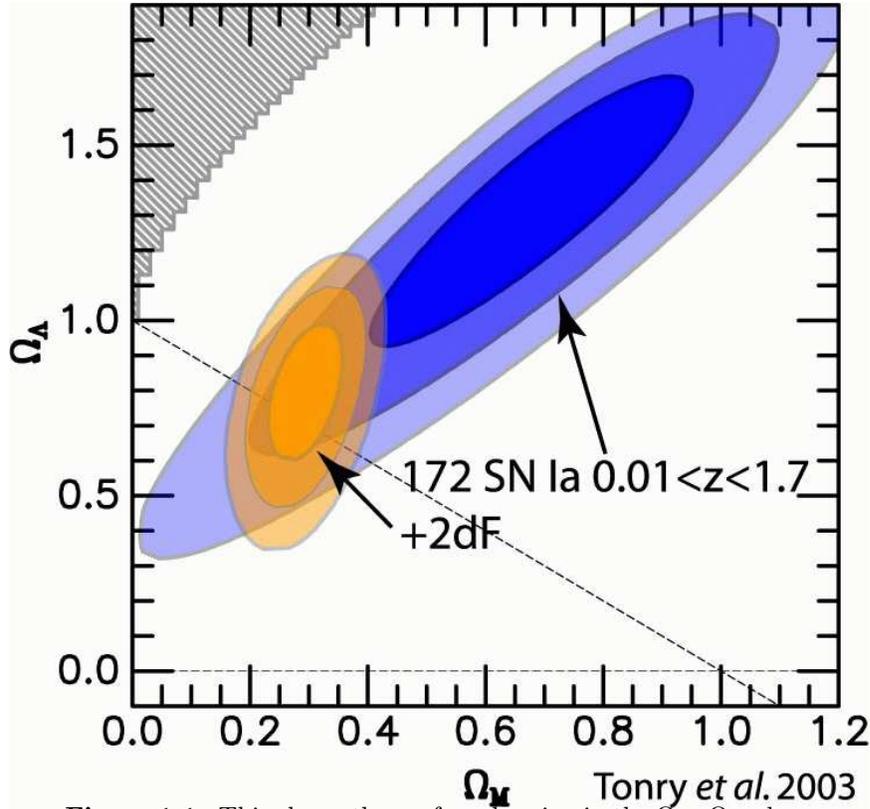

**Figure 1.1:** This shows the preferred region in the $\Omega_m$–$\Omega_\Lambda$ plane from a study of 172 supernovae, and also how the constraints tighten when the 2dF galaxy redshift survey power spectrum is added as an additional constraint. [Reproduced with permission from Tonry *et al.* [16]. ]

Two major studies, the 'Supernova Cosmology Project' and the 'High-$z$ Supernova Search Team', found evidence for an accelerating Universe [15], interpreted as due to a cosmological constant, or to a more general 'dark energy' component. Recent results obtained by Tonry *et al.* [16] are shown in Fig. 1.1 (see also Ref. 17). The SNe Ia data alone can only constrain a combination of $\Omega_m$ and $\Omega_\Lambda$. When combined with the CMB data (which indicates flatness, *i.e.*, $\Omega_m + \Omega_\Lambda \approx 1$), the best-fit values are $\Omega_m \approx 0.3$ and $\Omega_\Lambda \approx 0.7$. Future experiments will aim to set constraints on the cosmic equation of state $w(z)$. However, given the integral relation between the luminosity distance and $w(z)$, it is not straightforward to recover $w(z)$ (*e.g.*, Ref. 18).

### 1.3.3. *Cosmic microwave background*:

The physics of the cosmic microwave background (CMB) is described in detail by Scott and Smoot in this volume. Before recombination, the baryons and photons are tightly coupled, and the perturbations oscillate in the potential wells generated primarily by the dark matter perturbations. After decoupling, the baryons are free to collapse into those potential wells. The CMB carries a record of conditions at the time of decoupling, often called primary anisotropies. In addition, it is affected by various processes as it propagates towards us, including the effect of a time-varying gravitational potential (the integrated Sachs-Wolfe effect), gravitational lensing, and scattering from ionized gas at low redshift.

The primary anisotropies, the integrated Sachs-Wolfe effect, and scattering from a homogeneous distribution of ionized gas, can all be calculated using linear perturbation theory, a widely-used implementation being the CMBFAST code of Seljak and Zaldarriaga [5]. Gravitational lensing is also calculated in this code. Secondary effects such as inhomogeneities in the re-ionization process, and scattering from gravitationally-collapsed gas (the Sunyaev-Zel'dovich effect), require more complicated, and more uncertain, calculations.

The upshot is that the detailed pattern of anisotropies, quantified, for instance, by the angular power spectrum $C_\ell$, depends on all of the cosmological parameters. In a typical cosmology, the anisotropy power spectrum [usually plotted as $\ell(\ell+1)C_\ell$] features a flat plateau at large angular scales (small $\ell$), followed by a series of oscillatory features at higher angular scales, the first and most prominent being at around one degree ($\ell \simeq 200$). These features, known as acoustic peaks, represent the oscillations of the photon-baryon fluid around the time of decoupling. Some features can be closely related to specific parameters—for instance, the location of the first peak probes the spatial geometry, while the relative heights of the peaks probes the baryon density—but many other parameters combine to determine the overall shape.

The WMAP experiment [1] has provided the most accurate results to date on the spectrum of CMB fluctuations [19], with a precision determination of the temperature power spectrum up to $\ell \simeq 900$, shown in Fig. 1.2, and the first detailed measurement of the correlation spectrum between temperature and polarization [20] (the correlation having first been detected by DASI [21]) . These are consistent with models based on the parameters we have described, and provide quite accurate determinations of many of them [7]. In this subsection, we will refer to results from WMAP alone, with the following section combining those with other observations. We note that as the parameter fitting is done in a multi-parameter space, one has to assume a 'prior' range for each of the parameters (*e.g.*, Hubble constant $0.5 < h < 1$), and there may be some dependence on these assumed priors.

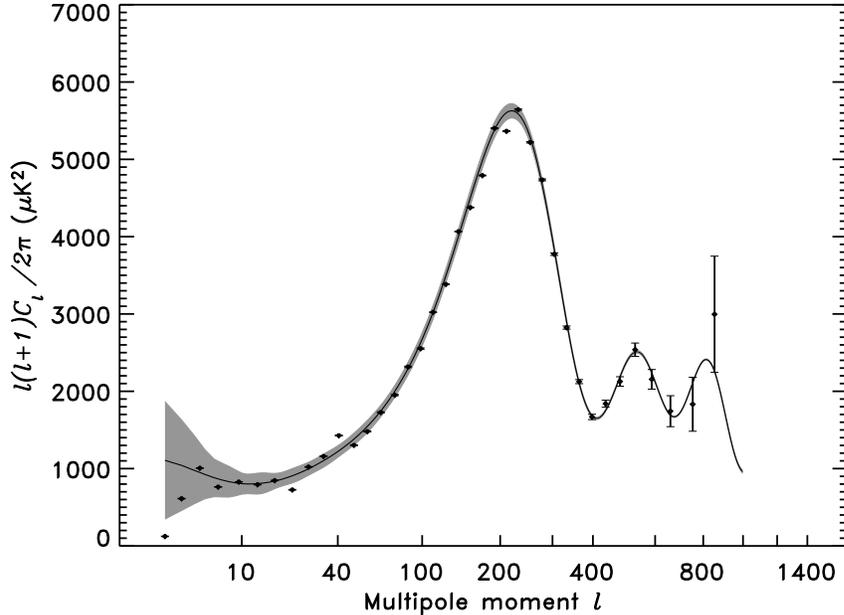

**Figure 1.2:** The angular power spectrum of the cosmic microwave background as measured by the WMAP satellite. The solid line shows the prediction from the best-fitting $\Lambda$CDM model [7]. The error bars on the data points (which are tiny for most of them) indicate the observational errors, while the shaded region indicates the statistical uncertainty from being able to observe only one microwave sky, known as cosmic variance, which is the dominant uncertainty on large angular scales. [Figure courtesy NASA/WMAP Science Team.]

WMAP provides an exquisite measurement of the location of the first acoustic peak, which directly probes the spatial geometry and yields a total density $\Omega_{\rm tot} \equiv \sum \Omega_i + \Omega_\Lambda$ of

$$\Omega_{\rm tot} = 1.02 \pm 0.02 \,, \tag{1.14}$$

consistent with spatial flatness and completely excluding significantly curved Universes (this result does however assume a fairly strong prior on the Hubble parameter from other measurements; WMAP alone constrains it only weakly, and allows significantly closed Universes if $h$ is small, e.g. $\Omega_{\rm tot} = 1.3$ for $h = 0.3$). It also gives a precision measurement of the age of the Universe. It gives a baryon density consistent with that coming from nucleosynthesis, and affirms the need for both dark matter and dark energy if the data are to be explained. For the spectral index of density perturbations, WMAP alone is consistent with a power-law spectrum, with spectral index $n = 0.99 \pm 0.04$, and in particular with a scale-invariant initial spectrum $n = 1$. It shows no evidence for dynamics of the dark energy, being consistent with a pure cosmological constant ($w = -1$).

One of the most interesting results, driven primarily by detection of large-angle polarization-temperature correlations, is the discovery of a high optical depth to re-ionization, $\tau \sim 0.17$, which roughly corresponds to a re-ionization redshift $z_{\mathrm{ion}} \sim 17$. This was higher than expected, though it appears it can be accommodated in models for development of the first structures which provide the ionizing flux.

In addition to WMAP, useful information comes from measurements of the CMB on small angular scales by, amongst others, the ACBAR and CBI experiments. Further, in 2002 the DASI experiment made the first measurement of the polarization anisotropies [21], again consistent with the standard cosmology, though not with sufficient accuracy to provide detailed constraints.

### 1.3.4. *Galaxy clustering*:

The power spectrum of density perturbations depends on the nature of the dark matter. Within the Cold Dark Matter model, the shape of the power spectrum depends primarily on the primordial power spectrum and on the combination $\Omega_{\mathrm{m}}h$ which determines the horizon scale at matter-radiation equality, with a subdominant dependence on the baryon density. The matter distribution is most easily probed by observing the galaxy distribution, but this must be done with care as the galaxies do not perfectly trace the dark matter distribution. Rather, they are a 'biased' tracer of the dark matter. The need to allow for such bias is emphasized by the observation that different types of galaxies show bias with respect to each other. Further, the observed 3D galaxy distribution is in redshift space, *i.e.*, the observed redshift is the sum of the Hubble expansion and the line-of-sight peculiar velocity, leading to linear and non-linear dynamical effects which also depend on the cosmological parameters. On the largest length scales, the galaxies are expected to trace the location of the dark matter, except for a constant multiplier $b$ to the power spectrum, known as the linear bias parameter. On scales smaller than 20 $h^{-1}$ Mpc or so, the clustering pattern is 'squashed' in the radial direction due to coherent infall, which depends on the parameter $\beta \equiv \Omega_{\mathrm{m}}^{0.6}/b$ (on these shorter scales, more complicated forms of biasing are not excluded by the data). On scales of a few $h^{-1}$ Mpc, there is an effect of elongation along the line of sight (colloquially known as the 'finger of God' effect) which depends on the galaxy velocity dispersion $\sigma_{\mathrm{p}}$.

**1.3.4.1.**   *The galaxy power spectrum:*

The 2-degree Field (2dF) Galaxy Redshift Survey is now complete and publicly available, with nearly 230,000 redshifts.** Analyses of a subset of the full data (containing 160,000 redshifts) measured the power spectrum for $k > 0.02\,h\,\mathrm{Mpc}^{-1}$ with $\sim 10\%$ accuracy, shown in Fig. 1.3. The measured power spectrum is well fit by a CDM model with $\Omega_{\mathrm{m}}h = 0.18 \pm 0.02$, and a baryon fraction $\Omega_{\mathrm{b}}/\Omega_{\mathrm{m}} = 0.17 \pm 0.06$ [22]. The pattern of the galaxy clustering in redshift space is fitted by $\beta = 0.49 \pm 0.09$ and velocity dispersion $\sigma_{\mathrm{p}} = 506 \pm 52\,\mathrm{km\,s}^{-1}$ [23]; note that the two are strongly correlated. Combination of the 2dF data with the CMB indicates $b \sim 1$, in agreement with a 2dF-alone analysis of higher-order clustering statistics. Results for these parameters also depend on the length scale over which a fit is done, and the selection of the objects by luminosity, spectral type, or color. In particular, on scales smaller than $10\ h^{-1}\mathrm{Mpc}$, different galaxy types are clustered differently. This 'biasing' introduces a systematic effect on the determination of cosmological parameters from redshift surveys. Prior knowledge from simulations of galaxy formation could help, but is model-dependent. We note that the present-epoch power spectrum is not sensitive to dark energy, so it is mainly a probe of the matter density.

The Sloan Digital Sky Survey (SDSS) is a project to image a quarter of the sky, and to obtain spectra of galaxies and quasars selected from the imaging data.†† A maximum likelihood analysis of early SDSS data by Szalay *et al.* [24] used the projected distribution of galaxies in a redshift bin around $z = 0.33$ to find $\Omega_{\mathrm{m}}h = 0.18 \pm 0.04$, assuming a flat $\Lambda$CDM model with $\Omega_{\mathrm{m}} = 1 - \Omega_{\Lambda} = 0.3$. The power spectrum of the latest version of SDSS redshift survey was published as this article was being finalized [25].

**1.3.4.2.**   *Limits on neutrino mass from 2dFGRS:*

Large-scale structure data can put an upper limit on the ratio $\Omega_{\nu}/\Omega_{\mathrm{m}}$ due to the neutrino 'free streaming' effect [26]. By comparing the 2dF galaxy power spectrum with a four-component model (baryons, cold dark matter, a cosmological constant, and massive neutrinos), it was estimated that $\Omega_{\nu}/\Omega_{\mathrm{m}} < 0.13$ (95% confidence limit), giving $\Omega_{\nu} < 0.04$ if a concordance prior of $\Omega_{\mathrm{m}} = 0.3$ is imposed. The latter corresponds to an upper limit of about 2 eV on the total neutrino mass, assuming a prior of $h \approx 0.7$ [27]. The above analysis assumes that the primordial power spectrum is adiabatic, scale-invariant and Gaussian. Potential systematic effects include biasing of the galaxy distribution and non-linearities of the power spectrum. Additional cosmological data sets bring down this upper limit by a factor of two [28]. The analysis of WMAP+2dFGRS [7] derived $\Omega_{\nu}h^2 < 0.0067$ (95% CL).

Laboratory limits on absolute neutrino masses from tritium beta decay and especially from neutrinoless double-beta decay should,

---

** See `http://www.mso.anu.edu.au/2dFGRS`

†† See `http://www.sdss.org`

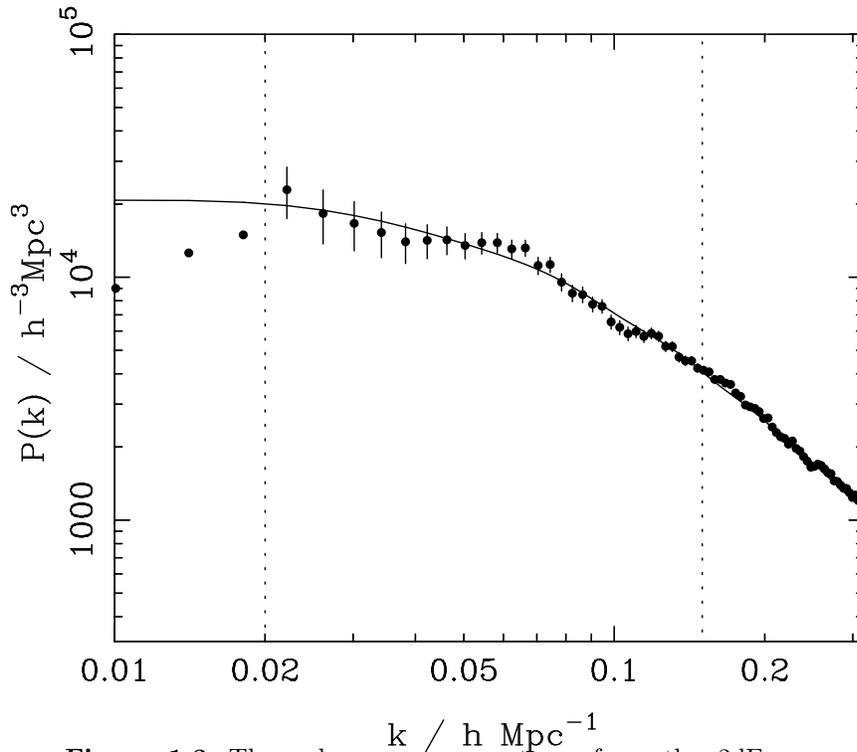

**Figure 1.3:** The galaxy power spectrum from the 2dF galaxy redshift survey as derived in Ref. 22. This plot shows $P(k) \propto \Delta^2(k)/k^3$, but with distances measured in redshift space and convolved with the survey geometry. The solid line shows a linear-theory $\Lambda$CDM fit (also convolved with the survey geometry) with $\Omega_m h = 0.2$, $\Omega_b/\Omega_m = 0.15$, $h = 0.7$ and $n = 1$. Only the range $0.02\,h\,{\rm Mpc}^{-1} < k < 0.15\,h\,{\rm Mpc}^{-1}$, where perturbations are in the linear regime, was used to obtain that best fit. The error bars are correlated, but with known covariances. [Figure provided by Will Percival; see also Ref. 22.]

within the next decade, push down towards (or perhaps even beyond) the 0.1 eV level that has cosmological significance.

### 1.3.5. *Clusters of galaxies*:

A cluster of galaxies is a large collection of galaxies held together by their mutual gravitational attraction. The largest ones are around $10^{15}$ solar masses, and are the largest gravitationally-bound structures in the Universe. Even at the present epoch they are relatively rare, with only a few percent of galaxies being in clusters. They provide various ways to study the cosmological parameters; here we discuss constraints from the measurements of the cluster number density and the baryon fraction in clusters.

**1.3.5.1.    *Cluster number density:*** The first objects of a given kind form at the rare high peaks of the density distribution, and if the primordial density perturbations are Gaussian-distributed, their number density is exponentially sensitive to the size of the perturbations, and hence can strongly constrain it. Clusters are an ideal application in the present Universe. They are usually used to constrain the amplitude $\sigma_8$, as a box of side $8\,h^{-1}\,\mathrm{Mpc}$ contains about the right amount of material to form a cluster. The most useful observations at present are of X-ray emission from hot gas lying within the cluster, whose temperature is typically a few keV, and which can be used to estimate the mass of the cluster. A theoretical prediction for the mass function of clusters can come either from semi-analytic arguments or from numerical simulations. At present, the main uncertainty is the relation between the observed gas temperature and the cluster mass, despite extensive study using simulations. A recent analysis [29] gives

$$\sigma_8 = 0.78^{+0.30}_{-0.06} \qquad (95\%\mathrm{CL}) \qquad (1.15)$$

for $\Omega_{\mathrm{m}} = 0.35$, with highly non-Gaussian error bars, but different authors still find a spread of values. Scaling to lower $\Omega_{\mathrm{m}}$ increases $\sigma_8$ somewhat, and the result above is consistent with values predicted in cosmologies compatible with WMAP.

The same approach can be adopted at high redshift (which for clusters means redshifts approaching one) to attempt to measure $\sigma_8$ at an earlier epoch. The evolution of $\sigma_8$ is primarily driven by the value of the matter density $\Omega_{\mathrm{m}}$, with a sub-dominant dependence on the dark energy density. It is generally recognized that such analyses favor a low matter density, though there is not complete consensus on this, and at present this technique for constraining the density is not competitive with the CMB.

**1.3.5.2.    *Cluster baryon fraction:*** If clusters are representative of the mass distribution in the Universe, the fraction of the mass in baryons to the overall mass distribution would be $f_{\mathrm{b}} = \Omega_{\mathrm{b}}/\Omega_{\mathrm{m}}$. If $\Omega_{\mathrm{b}}$, the baryon density parameter, can be inferred from the primordial nucleosynthesis abundance of the light elements, the cluster baryon fraction $f_{\mathrm{b}}$ can then be used to constrain $\Omega_{\mathrm{m}}$ and $h$ (*e.g.*, Ref. 30). The baryons in clusters are primarily in the form of X-ray-emitting gas that falls into the cluster, and secondarily in the form of stellar baryonic mass. Hence, the baryon fraction in clusters is estimated to be

$$f_{\mathrm{b}} = \frac{\Omega_{\mathrm{b}}}{\Omega_{\mathrm{m}}} \simeq f_{\mathrm{gas}} + f_{\mathrm{gal}}, \qquad (1.16)$$

where $f_{\mathrm{b}} = M_{\mathrm{b}}/M_{\mathrm{grav}}$, $f_{\mathrm{gas}} = M_{\mathrm{gas}}/M_{\mathrm{grav}}$, $f_{\mathrm{gal}} = M_{\mathrm{gal}}/M_{\mathrm{grav}}$, and $M_{\mathrm{grav}}$ is the total gravitating mass.

This can be used to obtain an approximate relation between $\Omega_{\mathrm{m}}$ and $h$:

$$\Omega_{\mathrm{m}} = \frac{\Omega_{\mathrm{b}}}{f_{\mathrm{gas}} + f_{\mathrm{gal}}} \simeq \frac{\Omega_{\mathrm{b}}}{0.08h^{-1.5} + 0.01h^{-1}}. \qquad (1.17)$$

Big Bang Nucleosynthesis gives $\Omega_b h^2 \approx 0.02$, allowing the above relation to be approximated as $\Omega_m h^{0.5} \approx 0.25$ (*e.g.*, Ref. 31). For example, Allen *et al.* [32] derived a density parameter consistent with $\Omega_m = 0.3$ from Chandra observations.

### 1.3.6. *Clustering in the inter-galactic medium*:

It is commonly assumed, based on hydrodynamic simulations, that the neutral hydrogen in the inter-galactic medium (IGM) can be related to the underlying mass distribution. It is then possible to estimate the matter power spectrum on scales of a few megaparsecs from the absorption observed in quasar spectra, the so-called Lyman-alpha forest. The usual procedure is to measure the power spectrum of the transmitted flux, and then to infer the mass power spectrum. Photo-ionization heating by the ultraviolet background radiation and adiabatic cooling by the expansion of the Universe combine to give a simple power-law relation between the gas temperature and the baryon density. It also follows that there is a power-law relation between the optical depth $\tau$ and $\rho_b$. Therefore, the observed flux $F = \exp(-\tau)$ is strongly correlated with $\rho_b$, which itself traces the mass density. The matter and flux power-spectra can be related by

$$P_m(k) = b^2(k)\ P_F(k)\ ,\qquad(1.18)$$

where $b(k)$ is a bias function which is calibrated from simulations. Croft *et al.* [33] derived cosmological parameters from Keck Telescope observations of the Lyman-alpha forest at redshifts $z = 2 - 4$. Their derived power spectrum corresponds to that of a CDM model, which is in good agreement with the 2dF galaxy power spectrum. A recent study using VLT spectra [34] agrees with the flux power spectrum of Ref. 33.

This method depends on various assumptions. Seljak *et al.* [35] pointed out that errors are sensitive to the range of cosmological parameters explored in the simulations, and the treatment of the mean transmitted flux. Combination of the Lyman-alpha data with WMAP suggested deviation from the scale-invariant $n = 1$ power spectrum [7,6], but Seljak *et al.* [35] have argued that the combined data set is still compatible with $n = 1$ model.

### 1.3.7. *Gravitational lensing*:

Images of background galaxies get distorted due to the gravitational effect of mass fluctuations along the line of sight. Deep gravitational potential wells such as galaxy clusters generate 'strong lensing', *i.e.*, arcs and arclets, while more moderate fluctuations give rise to 'weak lensing'. Weak lensing is now widely used to measure the mass power spectrum in random regions of the sky (see Ref. 36 for recent reviews). As the signal is weak, the CCD frame of deformed galaxy shapes ('shear map') is analyzed statistically to measure the power spectrum, higher moments, and cosmological parameters.

The shear measurements are mainly sensitive to the combination of $\Omega_m$ and the amplitude $\sigma_8$. There are various systematic effects in the interpretation of weak lensing, *e.g.*, due to atmospheric distortions during observations, the redshift distribution of the background galaxies, intrinsic correlation of galaxy shapes, and non-linear modeling uncertainties. Hoekstra *et al.* [37] derived the result $\sigma_8 \Omega_m^{0.52} = 0.46^{+0.05}_{-0.07}$ (95% confidence level), assuming a $\Lambda$CDM model. Other recent results are summarized in Ref. 36. For a $\Omega_m = 0.3, \Omega_\Lambda = 0.7$ cosmology, different groups derived normalizations $\sigma_8$ over a wide range, indicating that the systematic errors are still larger than some of the quoted error bars.

**1.3.8.** *Peculiar velocities*:

Deviations from the Hubble flow directly probe the mass fluctuations in the Universe, and hence provide a powerful probe of the dark matter. Peculiar velocities are deduced from the difference between the redshift and the distance of a galaxy. The observational difficulty is in accurately measuring distances to galaxies. Even the best distance indicators (*e.g.*, the Tully-Fisher relation) give an error of 15% per galaxy, hence limiting the application of the method at large distances. Peculiar velocities are mainly sensitive to $\Omega_m$, not to $\Omega_\Lambda$ or quintessence. Extensive analyses in the early 1990s (*e.g.*, Ref. 38) suggested a value of $\Omega_m$ close to unity. A more recent analysis [39], which takes into account non-linear corrections, gives $\sigma_8 \Omega_m^{0.6} = 0.49 \pm 0.06$ and $\sigma_8 \Omega_m^{0.6} = 0.63 \pm 0.08$ (90% errors) for two independent data sets. While at present cosmological parameters derived from peculiar velocities are strongly affected by random and systematic errors, a new generation of surveys may improve their accuracy. Two promising approaches are the 6dF near-infrared survey of 15,000 peculiar velocities[‡‡] and the kinematic Sunyaev-Zel'dovich effect.

## 1.4. Bringing observations together

Although it contains two ingredients—dark matter and dark energy—which have not yet been verified by laboratory experiments, the $\Lambda$CDM model is almost universally accepted by cosmologists as the best description of present data. The basic ingredients are given by the parameters listed in Sec. 1.1.4, with approximate values of some of the key parameters being $\Omega_b \approx 0.04$, $\Omega_{dm} \approx 0.26$, $\Omega_\Lambda \approx 0.70$, and a Hubble constant $h \approx 0.7$. The spatial geometry is very close to flat (and often assumed to be precisely flat), and the initial perturbations Gaussian, adiabatic, and nearly scale-invariant.

The most powerful single experiment is WMAP, which on its own supports all these main tenets. Values for some parameters, as given in Spergel *et al.* [7], are reproduced in Table 1.2. This model presumes a flat Universe, and so $\Omega_\Lambda$ is a derived quantity in this analysis, with best-fit value $\Omega_\Lambda = 0.73$.

However, to obtain the most powerful constraints, other data sets need to be considered in addition to WMAP. A standard data compilation unites WMAP with shorter-scale CMB measurements from CBI and ACBAR, and the galaxy power spectrum from the 2dF survey. In our opinion, this combination of datasets offers the most reliable set of constraints at present. In addition, it is possible to add the Lyman-alpha forest power spectrum data, but this has proven more controversial as the interpretation of such data has not reached a secure level.

---

[‡‡] See http://www.mso.anu.edu.au/6dFGS/

**Table 1.2:** Parameter constraints reproduced from Spergel *et al.* [7], both from WMAP alone and from the preferred data compilation of WMAP+CBI+ACBAR (known as WMAPext) plus 2dFGRS. The first two columns assume a power-law initial spectrum, while the third allows a running of the spectral index (in this case $n$ is defined at a particular scale, and its value cannot be directly compared with the power-law case). Spatial flatness is assumed in the parameter fit. The parameter $A$ is a measure of the perturbation amplitude; see Ref. 7 for details. Uncertainties are shown at one sigma, and caution is needed in extrapolating them to higher significance levels due to non-Gaussian likelihoods and assumed priors.

|  | WMAP alone power-law | WMAPext + 2dFGRS power-law | WMAPext + 2dFGRS running |
|---|---|---|---|
| $\Omega_{\mathrm{m}}h^2$ | $0.14 \pm 0.02$ | $0.134 \pm 0.006$ | $0.136 \pm 0.009$ |
| $\Omega_{\mathrm{b}}h^2$ | $0.024 \pm 0.001$ | $0.023 \pm 0.001$ | $0.022 \pm 0.001$ |
| $h$ | $0.72 \pm 0.05$ | $0.73 \pm 0.03$ | $0.71 \pm 0.04$ |
| $n$ | $0.99 \pm 0.04$ | $0.97 \pm 0.03$ | $0.93^{+0.04}_{-0.05}$ |
| $\tau$ | $0.17^{+0.08}_{-0.07}$ | $0.15 \pm 0.07$ | $0.17 \pm 0.06$ |
| $A$ | $0.9 \pm 0.1$ | $0.8 \pm 0.1$ | $0.84 \pm 0.09$ |
| $dn/d\ln k$ | - | - | $-0.031^{+0.023}_{-0.025}$ |

Using the extended data set without the Lyman-alpha constraints produces no surprises; as compared to WMAP alone, the best-fit values move around a little within the uncertainties, and the error bars improve somewhat, as seen in Table 1.2. In this table we also show the effect of allowing the spectral index to vary with scale ('running'): the running is found to be consistent with zero and there are small drifts in the values and uncertainties of the other parameters.¶

However, inclusion of the Lyman-alpha data suggests a more radical development, with the running weakly detected at around 95% confidence, the spectral index making a transition from $n > 1$ on large scales to $n < 1$ on small scales [7,6]. The significance of this measurement is not high, and the result rather unexpected on theoretical grounds (it suggests that the power spectrum has a maximum which just happens to lie in the rather narrow range of scales that observations are able to probe, and the running is much larger than in typical inflation models giving a spectral index close to one). In our view it is premature to read much significance into this observation, though if true, it should rapidly be firmed up by new data.

¶ As we were finalizing this article, an analysis of WMAP combined with the SDSS galaxy power spectrum appeared [40], giving results in good agreement with those discussed here.

The baryon density $\Omega_b$ is now measured with quite high accuracy from the CMB and large-scale structure, and shows reasonable agreement with the determination from big bang nucleosynthesis; Fields and Sarkar in this volume quote the range $0.012 \geq \Omega_b h^2 \geq 0.025$. Given the sensitivity of the measurement, it is important to note that it has significant dependence on both the datasets and parameter sets chosen, as seen in Table 1.2.

While $\Omega_\Lambda$ is measured to be non-zero with very high confidence, there is no evidence of evolution of the dark energy density. The WMAP team find the limit $w < -0.78$ at 95% confidence from a compilation of data including SNe Ia data, where they impose a prior $w \geq -1$, with the cosmological constant case $w = -1$ giving an excellent fit to the data.

As far as inflation is concerned, the data provide good news and bad news. The good news is that WMAP supports all the main predictions of the simplest inflation models: spatial flatness and adiabatic, Gaussian, nearly scale-invariant density perturbations. But it is disappointing that there is no sign of primordial gravitational waves, with WMAP providing only a weak upper limit $r < 0.53$ at 95% confidence [6]  (this assumes no running, and weakens significantly if running is allowed), and especially that no convincing deviations from scale-invariance have been seen. It is perfectly possible for inflation models to give $n \simeq 1$ and $r \simeq 0$, but in that limit, the observations give no clues as to the dynamical processes driving inflation. Tests have been made for various types of non-Gaussianity, a particular example being a parameter $f_{NL}$ which measures a quadratic contribution to the perturbations and is constrained to $-58 < f_{NL} < 134$ at 95% confidence [41]  (this looks weak, but prominent non-Gaussianity requires the product $f_{NL} \Delta_{\mathcal{R}}$ to be large, and $\Delta_{\mathcal{R}}$ is of order $10^{-5}$).

Two parameters which are still uncertain are $\Omega_m$ and $\sigma_8$ (see Figure 1.4 and Ref. 42). The value of $\Omega_m$ is beginning to be pinned down with some precision, with most observations indicating a value around 0.3, including the CMB anisotropies, the cluster number density, and gravitational lensing, though the latter two have a strong degeneracy with the amplitude of mass fluctuations $\sigma_8$. However, not all observations yet agree fully on this, for instance mass-to-light ratio measurements give $\Omega_m \approx 0.15$ [43],  and the fractional uncertainty remains significantly higher than one would like. Concerning $\sigma_8$, results from the cluster number density have varied quite a lot in recent years, spanning the range 0.6 to 1.0, primarily due to the uncertainties in the mass-temperature-luminosity relations used to connect the observables with theory. There is certainly scope for improving this calibration by comparison to mass measurements from strong gravitational lensing. The WMAP-alone measurements gives $\sigma_8 = 0.9 \pm 0.1$. However, this is not a direct constraint; WMAP only probes larger length scales, and the constraint comes from using WMAP to estimate all the parameters of the model needed to determine $\sigma_8$. As such, their constraint depends strongly on the assumed set of cosmological parameters being sufficient.

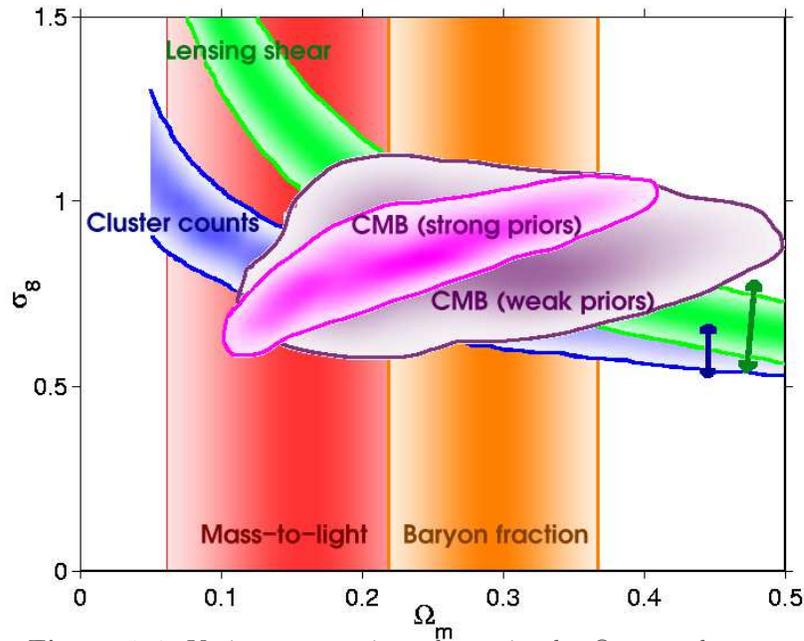

**Figure 1.4:** Various constraints shown in the $\Omega_m$–$\sigma_8$ plane. [Figure provided by Sarah Bridle; see also Ref. 42.]

One parameter which is surprisingly robust is the age of the Universe. There is a useful coincidence that for a flat Universe the position of the first peak is strongly correlated with the age of the Universe. The WMAP-only result is $13.4 \pm 0.3$ Gyr (assuming a flat Universe). This is in good agreement with the ages of the oldest globular clusters [44] and radioactive dating [45].

## 1.5. Outlook for the future

The concordance model is now well established, and there seems little room left for any dramatic revision of this paradigm. A measure of the strength of that statement is how difficult it has proven to formulate convincing alternatives. For example, one corner of parameter space that has been explored is the possibility of abandoning the dark energy, and instead considering a mixed dark matter model with $\Omega_m = 1$ and $\Omega_\nu = 0.2$. Such a model fits both the 2dF and WMAP data reasonably well, but only for a Hubble constant $h < 0.5$ [27,46]. However, this model is inconsistent with the HST key project value of $h$, the results from SNe Ia, cluster number density evolution, and baryon fraction in clusters.

Should there indeed be no major revision of the current paradigm, we can expect future developments to take one of two directions. Either the existing parameter set will continue to prove sufficient to explain the data, with the parameters subject to ever-tightening constraints, or it will become necessary to deploy new parameters. The latter outcome would be very much the more interesting, offering a route towards understanding new physical processes relevant to the cosmological evolution. There are many possibilities on offer for striking discoveries, for example:

- The cosmological effects of a neutrino mass may be unambiguously detected, shedding light on fundamental neutrino properties;

- Detection of deviations from scale-invariance in the initial perturbations would indicate dynamical processes during perturbation generation, for instance, by inflation;

- Detection of primordial non-Gaussianities would indicate that non-linear processes influence the perturbation generation mechanism;

- Detection of variation in the dark energy density (*i.e.*, $w \neq -1$) would provide much-needed experimental input into the question of the properties of the dark energy.

These provide more than enough motivation for continued efforts to test the cosmological model and improve its precision.

Over the coming years, there are a wide range of new observations, which will bring further precision to cosmological studies. Indeed, there are far too many for us to be able to mention them all here, and so we will just highlight a few areas.

The cosmic microwave background observations will improve in several directions. The new frontier is the study of polarization, first detected in 2002. Data are imminent from balloon-based experiments including Maxipol and Boomerang, and with WMAP continuing to take data, they should be able to measure a polarization spectrum, as well as improve measures of the temperature-polarization cross-correlation (which is easier to measure as the temperature anisotropies are much larger). Dedicated ground-based polarization experiments, such as CBI and QUEST, promise powerful measures of the polarization spectrum in the next few years, and may be able to separately detect the two modes of polarization. Another area of development is pushing accurate power spectrum measurements to smaller angular scales, typically achieved by interferometry, which should allow measurements of secondary anisotropy effects, such as the Sunyaev-Zel'dovich effect, whose detection has already been tentatively claimed by CBI. Finally, we mention the Planck satellite, due to launch in 2007, which will make high-precision all-sky maps of temperature and polarization, utilizing a very wide frequency range for observations to improve understanding of foreground contaminants, and to compile a large sample of clusters via the Sunyaev-Zel'dovich effect.

Concerning galaxy clustering, the Sloan Digital Sky Survey is well underway, and currently expected to yield around 600,000 galaxy redshifts covering one quarter of the sky. Large samples of galaxy positions at high redshifts ($z \sim 1$) will begin to be obtained, for instance, by the DEEP2 survey using the Keck telescopes, and the VIRMOS survey on the VLT. The 6dF survey aims to take high-quality redshift and peculiar velocity data for a large sample of nearby galaxies, and has already taken around 40,000 of the planned 170,000 redshifts.

Still awaiting final approval is the SNAP satellite, which seeks to carry out a survey for Type Ia supernovae out to redshifts approaching two, which should in particular be a powerful probe of the dark energy. With large samples, it may be possible to detect evolution of the dark energy density, thus measuring its equation of state. SNAP is also able to carry out a large weak gravitational lensing survey, complementing those becoming possible with large-format CCDs on ground-based telescopes. Before SNAP, the ESSENCE project will significantly increase the size of the SNe Ia dataset.

The development of the first precision cosmological model is a major achievement. However, it is important not to lose sight of the motivation for developing such a model, which is to understand the underlying physical processes at work governing the Universe's evolution. On that side, progress has been much less dramatic. For instance, there are many proposals for the nature of the dark matter, but no consensus as to which is correct. The nature of the dark energy remains a mystery. Even the baryon density, now measured to an accuracy of a few percent, lacks an underlying theory able to predict it even within orders of magnitude. Precision cosmology may have arrived, but at present many key questions remain unanswered.

*Acknowledgments:*

ARL was supported in part by the Leverhulme Trust. We thank Sarah Bridle and Jochen Weller for useful comments on this article, and OL thanks members of the Cambridge Leverhulme Quantitative Cosmology and 2dFGRS Teams for helpful discussions.